\documentclass[conference]{IEEEtran}
\IEEEoverridecommandlockouts
\usepackage[utf8]{inputenc}      % Ensure UTF-8 input support
\usepackage[T1]{fontenc}         % Ensure correct font encoding
\usepackage{amsmath,amssymb,amsfonts}
\usepackage{algorithmic}
\usepackage{graphicx}
\usepackage{textcomp}
\usepackage{xcolor}
\usepackage{comment}
\usepackage{cite}
\usepackage[hidelinks]{hyperref}
\usepackage{float}
\usepackage[caption=false,font=footnotesize]{subfig}

\newcommand{\unicamp}{\includegraphics[scale=0.025]{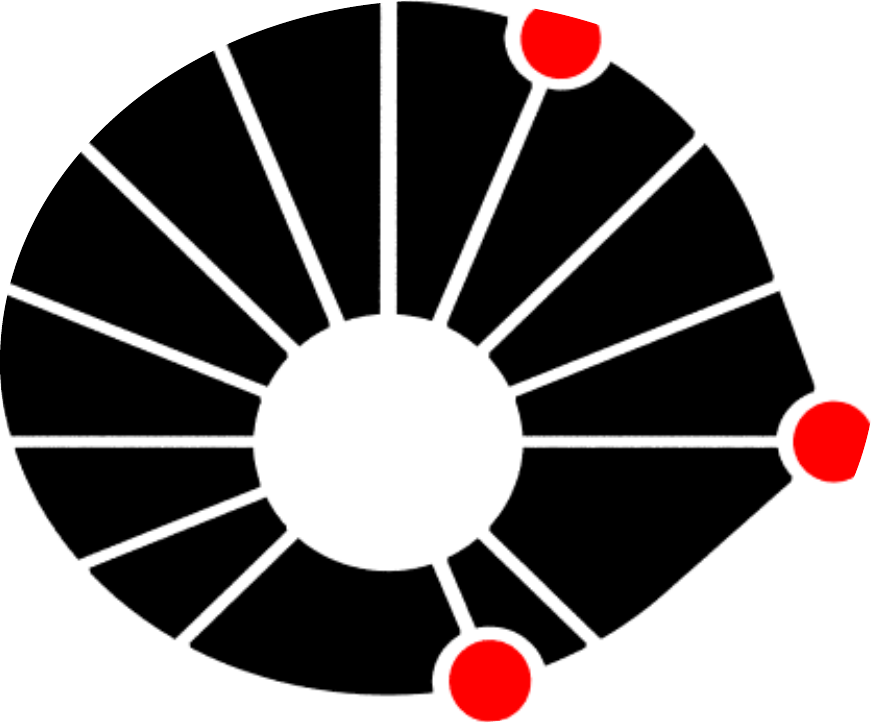}}    % must be .pdf/.png/.jpg
\newcommand{\fapesp}{\includegraphics[scale=0.035]{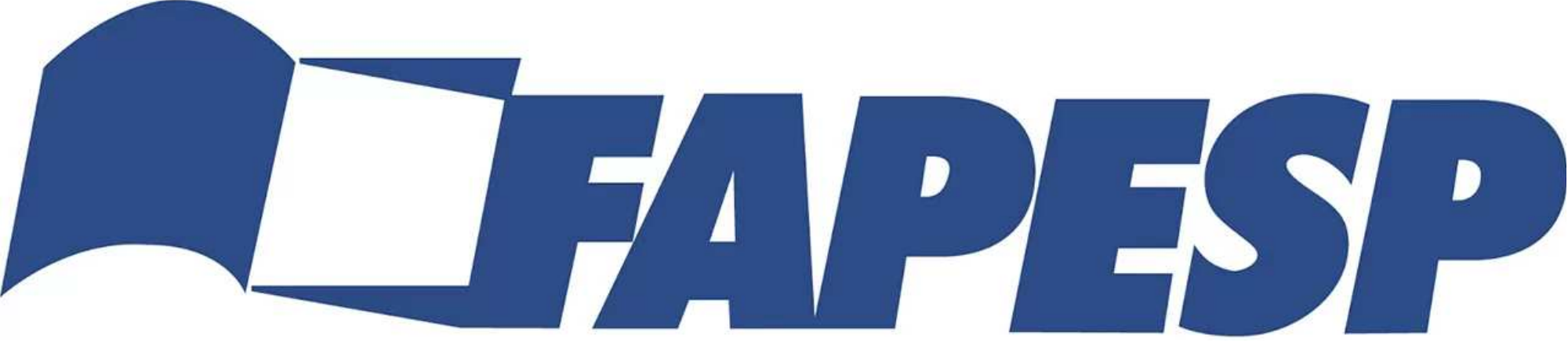}}    % same here
\newcommand{\smartness}{\includegraphics[scale=0.158]{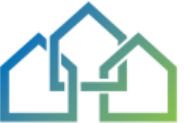}}
\newcommand{\ufscar}{\includegraphics[scale=0.05]{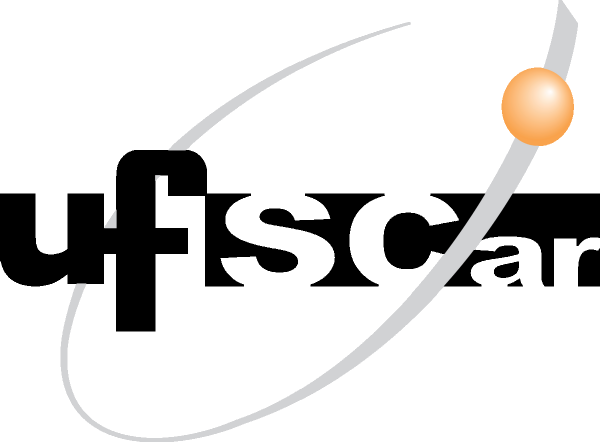}}

\def\BibTeX{{\rm B\kern-.05em{\sc i\kern-.025em b}\kern-.08em
    T\kern-.1667em\lower.7ex\hbox{E}\kern-.125emX}}

\begin{document}

\title{CGReplay: Capture and Replay of Cloud Gaming Traffic for QoE/QoS Assessment}
%CGReplay: Capture and Replay Cloud Gaming for QoE/QoS Assessment} 
%for QoE Assessment}

\author{
    \IEEEauthorblockN{Alireza Shirmarz~\ufscar, Ariel G. de Castro~\unicamp, Fabio L. Verdi~\ufscar, Christian E. Rothenberg~\unicamp}
    \IEEEauthorblockA{\ufscar~Universidade Federal de São Carlos~(UFSCar) - SP, Brazil
    \\\{ashirmarz, verdi\}@ufscar.br}
    \IEEEauthorblockA{\unicamp~Universidade Estadual de Campinas~(UNICAMP) - SP, Brazil
    \\a272319@dac.unicamp.br chesteve@dca.fee.unicamp.br}
}

\maketitle

\begin{abstract}
Cloud Gaming (CG) research faces challenges due to the unpredictability of game engines and restricted access to commercial platforms and their logs.
This creates major obstacles to conducting fair experimentation and evaluation.
\texttt{CGReplay} captures and replays player commands and the corresponding video frames in an ordered and synchronized action-reaction loop, ensuring reproducibility. %and fairnes.
It enables Quality of Experience/Service (QoE/QoS) assessment under varying network conditions and serves as a foundation for broader CG research. The code is publicly available for further development\footnote{\url{https://github.com/dcomp-leris/CGReplay.git}.}.
\end{abstract}


%\section{Key Networking Concepts}
\section{Introduction \& Motivation}
Cloud Gaming (CG) is rapidly growing as a popular entertainment medium, making both objective and subjective Quality of Experience (QoE) critical areas of research and industry focus~\cite{heo2024adrenaline, graff2024improving}. However, most mainstream CG platforms (\textit{e.g.}, Xbox Cloud Gaming, GeForce Now)
%-- such as Xbox Cloud Gaming, PlayStation Now, GeForce Now, and Amazon Luna --
are closed-source, hampering controlled research experiments and evaluation, restricting the exploration of innovative research directions in areas such as frame generation techniques, novel QoE metrics development, latency optimization strategies, and adaptive streaming algorithms tailored specifically for interactive gaming contexts. Additionally, the inherently non-deterministic nature of gameplay means that the same scene is never repeated, preventing direct comparison of video frames across different network conditions. Since a CG session involves uplink commands (player input) and downlink video frames (server response), their interactive behavior -- particularly, response time -- is essential to QoE. Therefore, a platform that can capture, synchronize, and replay these interactions is essential for reproducible testing and QoE assessment.

While open-source CG platforms such as GamingAnywhere\footnote{\url{https://github.com/chunying/gaminganywhere.git}.} and Moonlight Game Streaming\footnote{\url{https://github.com/moonlight-stream/moonlight-stream.github.io.git}.} exist, they also present challenges for QoE evaluation due to gameplay non-determinism -- where commands and video frames vary unpredictably. This variability makes it difficult to assess downlink video frame quality, uplink command accuracy, and their interactive influences under different network conditions. For generating CG traffic and evaluating its QoE, a platform is needed to capture CG over the Internet and replay the same uplink and downlink sequences. Such a solution would allow researchers to compare received video frames and evaluate how network conditions affect interactivity between commands and frames.

Experimental Cloud Gaming Platform (eCGP)~\cite{graff2024improving} aimed to achieve deterministic cloud gaming outputs by bypassing the non-determinism of game engines for Quality of Experience (QoS) evaluation. However, this approach does not fully support the automatic interactivity between uplink and downlink. Moreover, they fail to address real-world cloud gaming challenges, such as command and frame loss, which impact fidelity. Inspired by TCPReplay, we propose \texttt{CGReplay}\footnote{This is a complementary work supporting the full paper `\textit{In-Network AR/CG Traffic Classification Entirely Deployed in the Programmable Data Plane}` accepted at Netsoft 2025 main track.}, an open-source, configurable platform that captures and replays CG sessions at the application layer while preserving automatic interactivity. By synchronizing the timing and sequence of commands and frames, \texttt{CGReplay} allows researchers to conduct reproducible QoE experiments under various network conditions, ensuring higher fidelity to real-world cloud gaming environments.
\texttt{CGReplay} is organized into two main phases:
\begin{enumerate} \item \textbf{CG Capturing:} in this phase, the platform records uplink commands and downlink video frames along with their interaction, ensuring that the complete behavioral pattern of a cloud gaming session is preserved. 
\item \textbf{CG Replaying:} the recorded data is synchronized and replayed to mimic the original session, enabling reproducible testing under diverse network conditions. 
\end{enumerate}

For this demonstration, our implementation uses UDP, RTP, and SCReAM\footnote{\url{https://github.com/EricssonResearch/scream.git}.}, however, as an open-source platform, it can be extended for other streaming protocols such as RTP over QUIC (RoQ)~\cite{RoQ@ietf-avtcore-rtp-over-quic-13}. \texttt{CGReplay} can also be used as the platform for evaluation end development of other research CG adaptive rendering improvement ~\cite{heo2024adrenaline}, the adaptive encoding~\cite{salsify@2018} for CG and Generative interactive environments (Genie)~\cite{bruce2024genie}.

\section{Highlights of CGReplay}
%CGReplay is an open source, publicly available platform for capturing and replaying cloud gaming sessions, organized into three parts: capturing, replaying, and action/reaction interactivity and reliability. 
\texttt{CGReplay} is an open-source platform for capturing and replaying bi-directional cloud gaming sessions (Action/Reaction) based on real online CG interactions. It offers easy configuration via YAML files for each component, ensuring flexibility and simplicity. 
Its main contributions are structured into three key parts: capturing, replaying, and ensuring action/reaction interactivity and reliability. In the capturing phase, the platform records uplink commands and downlink video frames while extracting their sequential patterns. During replaying, these sequences are utilized by two agents: (a)~\textit{CG server} and (b)~\textit{CG player} running on Python-supported systems, to faithfully mimic the original session. Finally, the action/reaction interactivity module emulates user actions and server responses, with SCReAM (optional) integration for UDP-based congestion control to evaluate ECN-based signaling alongside traditional drop mechanisms.

\subsection{Capturing the Cloud Gaming}
\begin{figure}[!t]
  \centering
  \includegraphics[width=0.98\columnwidth]{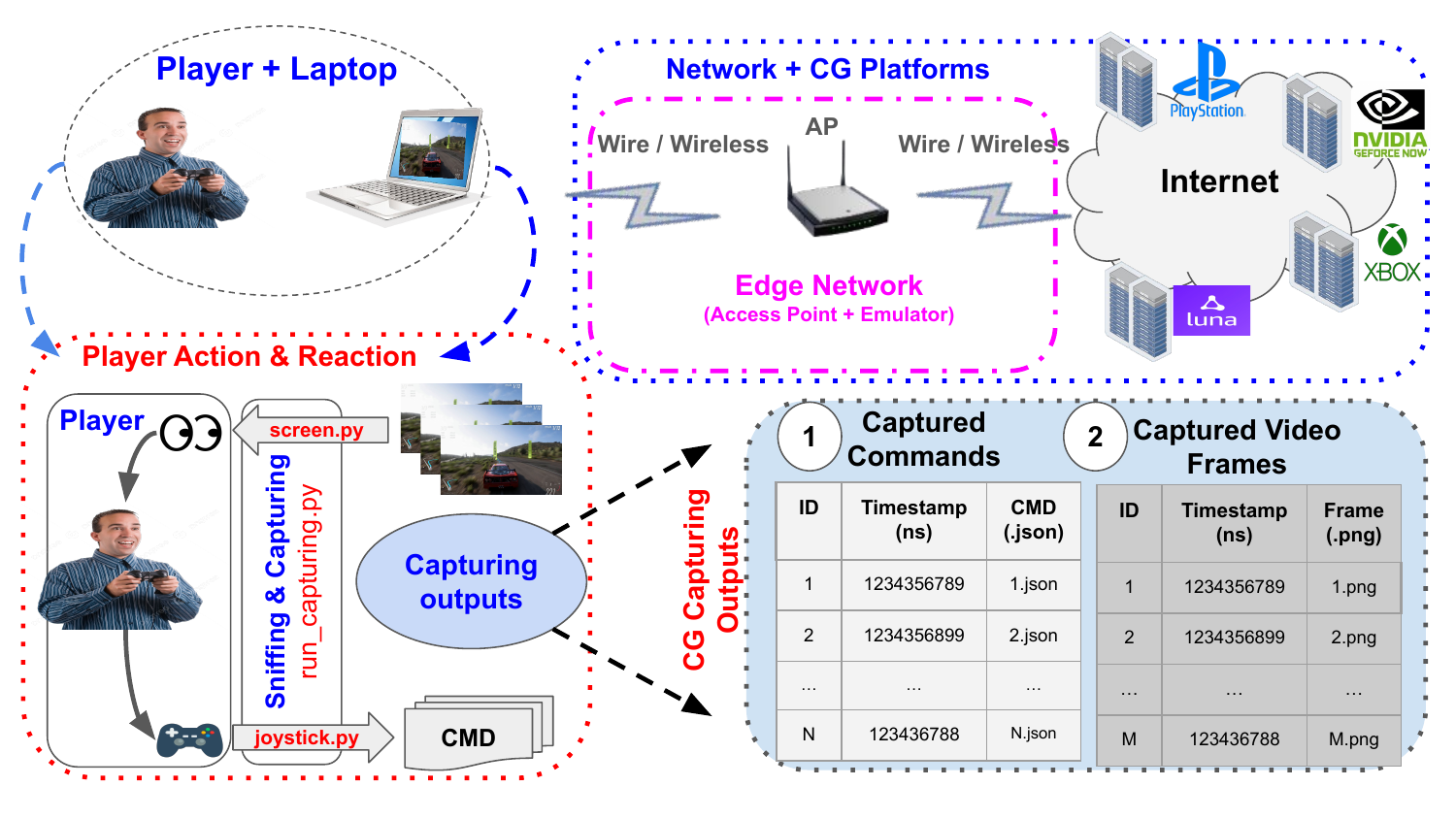} % or adjust width as needed
  \caption{Capturing online CG. The AP is connected to the Internet via a wired connection and connects to the laptop using Wi-Fi 6.}
  \label{fig:capture}
  \vspace{-5pt}
\end{figure}

%In the capture phase, we first select an appropriate network topology -- typically an Access Point (AP) connected to the Internet (or an edge network emulator). Next, we choose the cloud gaming platform (\textit{e.g.}, Xbox, PlayStation, or others) and a range of games (both free and commercial) to play. Finally, we execute the run-capturing.py script, which leverages screen.py and joystick.py to monitor I/O events, capturing both player commands and on-screen video (see Fig.~\ref{fig:capture}). This module runs solely on the player's laptop to capture the necessary logs. The output of this phase is a set of synchronized logs containing commands and video frames, complete with order and timestamps for subsequent replay.
In this phase, the module \texttt{\textit{run\_capturing.py}} captures both uplink commands (sniffed via the USB port) and downlink video frames (captured from the screen) at a configurable sampling rate of 30 frames per second. In our demonstration, a human player uses an Xbox cloud gaming platform to play three games -- Forza Horizon 5, Fortnite, and Mortal Kombat 11, as shown in Fig.~\ref{fig:capture}. This flexible setup accommodates various platforms and topologies (\textit{e.g.}, PlayStation, GeForce Now, Luna) and generates two synchronized files: one containing commands stored in a JSON file and the other containing video frames as PNG files, both annotated with IDs and timestamps. These logs are then used to extract a synchronized command/frame sequence for accurate sequential command and frame replay. Although the capturing module can record various inputs, in this study we focus on the joystick -- specifically, the Xbox Controller.

\subsection{Replaying the Cloud Gaming}
\begin{figure}[!t]
  \centering
  \includegraphics[width=0.99\columnwidth]{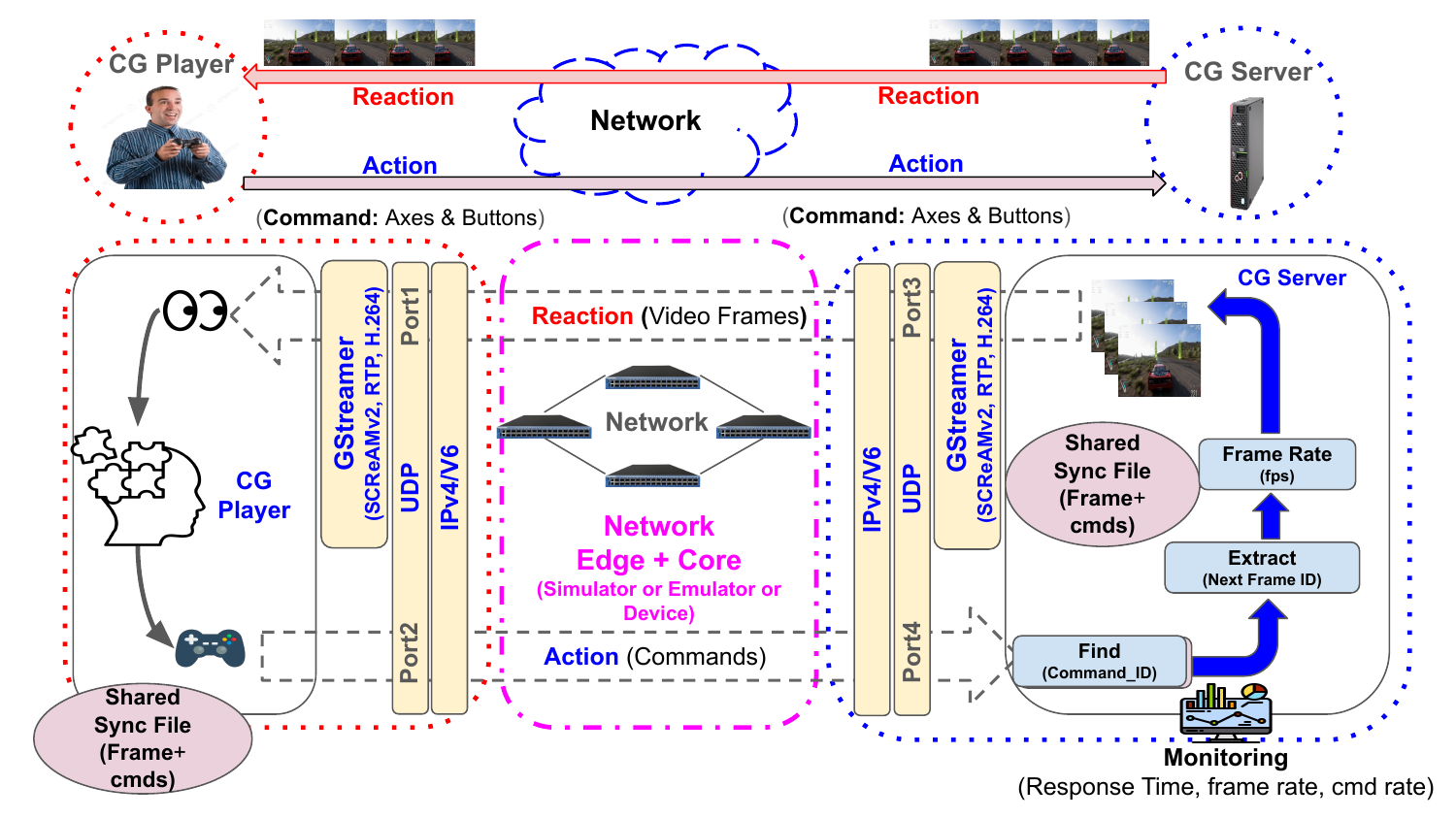} % or adjust width as needed
    \vspace{-10pt}
  \caption{Replaying CG traffic with CGReplay agents at server and player sides.}
  \label{fig:CGReplay}
  \vspace{-10pt}
\end{figure}

\texttt{CGReplay} comprises two Python agents -- (a)~\textit{CG server} and (b)~\textit{CG player} -- deployed on separate hosts that can be interconnected via physical devices or simulated/emulated networks (see Fig.~\ref{fig:CGReplay}). These modules use IPv4 and UDP for both uplink and downlink communications. Video streaming is implemented with GStreamer\footnote{\url{https://gstreamer.freedesktop.org/}.}, with RTP and H.264 encoding, with optional SCReAM integration as a scalable congestion control mechanism. The synchronized sequence of commands and frames ensures sequential replay between the \textit{CG player} and \textit{CG server}. During replay, the CG server streams frames sequentially until a command is needed. The CG player processes each frame, triggers the corresponding command, and sends it to the server. Streaming resumes upon receiving the expected command, ensuring proper interactivity.

% 
\begin{comment}
\begin{figure*}[htbp] %[!H]
  \centering
  \includegraphics[width=0.95\textwidth]{CGReplay.pdf}
  \caption{CGReplay is replaying CG session based on pre-captured CG, and it includes two agents: Player and Server. (1) CG player act with commands, and (2) CG server react with the video frames in response to the commands.}
  \label{fig:CGReplay}
\end{figure*}
\end{comment}

\begin{figure}[H]
  \centering
  \includegraphics[width=0.98\columnwidth]{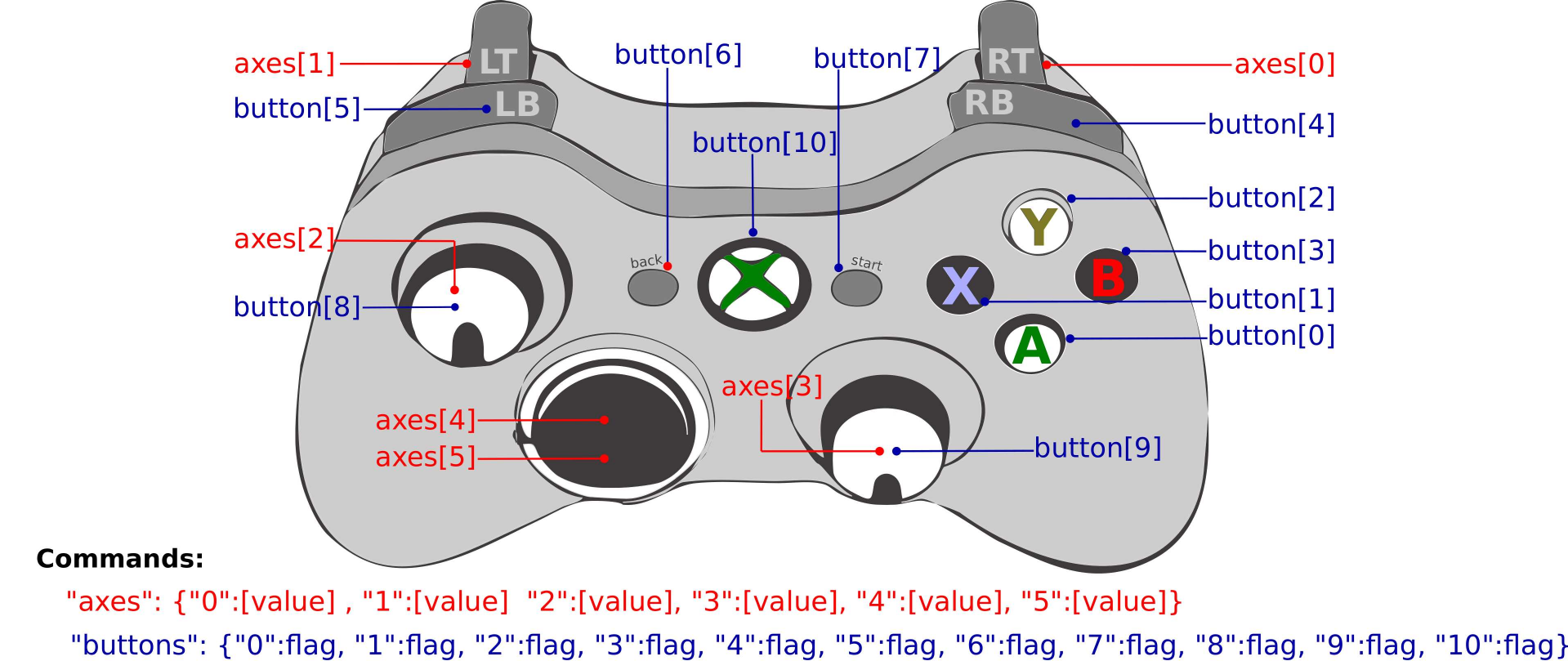} % or adjust width as needed
  \caption{Xbox Joystick Button Mapping with Corresponding Commands. The axes' continuous values range [-32767, 32767], and the button flag is 0 or 1.}
  \label{fig:joystick}
\end{figure}

\noindent\textbf{Game Video Frame:} video frames are stored as PNG files in the CG server's folder and sent in response to specific commands. Each frame is dynamically labeled with a unique ID and timestamp embedded as a QR code, allowing the CG player to verify frames during replay. \\
%Video frames are stored alongside the CG server agent in a dedicated folder as PNG files and are sent in response to specific commands. Each frame is augmented on-the-fly with a unique ID and timestamp embedded as a QR code, allowing the CG player module to examine and verify frames during replay.\\
\noindent\textbf{Commands:} commands are stored in JSON format with ``axes'' for joystick movements and ``button'' for presses (see Fig.~\ref{fig:joystick}). Each command includes a unique ID, timestamp, and ACK/NACK, ensuring synchronization and sequence order between the CG server and player.

%Commands are transmitted in JSON format with two fields: ``axes'' for analog joystick movements and ``button'' for binary button presses (see Fig.~\ref{fig:joystick}). Each command message is hashed and augmented with a header containing metadata for interactivity and synchronization. This header includes a unique ID, timestamp, Ack/Nack status, and the number of received frames, enabling the CG server to maintain sequence order and ensure precise command/frame synchronization.   

\subsection{Action/Reaction Interactivity \& Reliability Mechanism}
\begin{figure}[!t]
  \centering
  %\includesvg[width=0.8\columnwidth]{6_interactivity}
  \includegraphics[width=0.9\columnwidth]{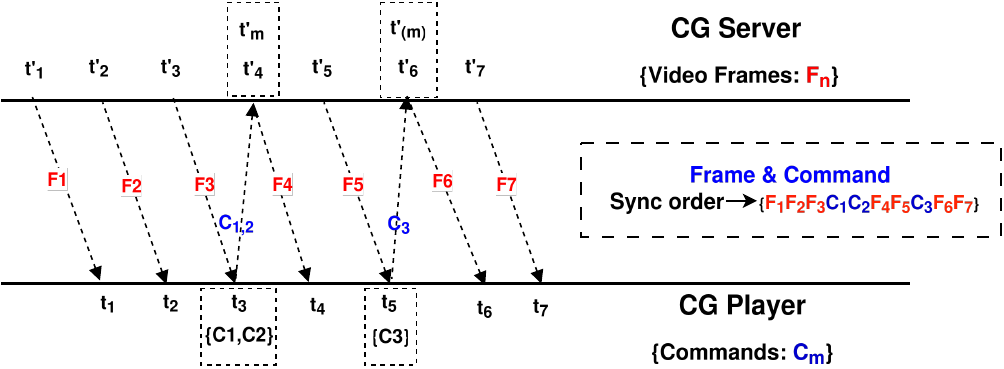} % or adjust width as needed
  \caption{Action \& Reaction Interactivity in CGReplay.}
  \label{fig:interactivity}
    \vspace{-10pt}
\end{figure}

In a real-time CG platform, the interactive loop works by having the player watch the current video frames, make decisions (\textit{e.g.}, pressing buttons), and then see updated frames influenced by those commands. To replicate this behavior automatically in \texttt{CGReplay}, the system uses a shared ``sync order'' of frames and commands, each with unique IDs ($F_{ID}$ for frames, $C_{ID}$ for commands). As shown in Fig.~\ref{fig:interactivity}, an example sync order might be  $\{F_1,F_2,F_3,C_1,C_2,F_4,F_5,C_3,F_6,F_7\}$. Both the \textit{CG server} and the \textit{CG player} compare incoming IDs against this sequence to ensure each action (command) is followed by its corresponding reaction (frame) at the correct time.
Meanwhile, on the \textit{CG server} side, frames are streamed sequentially, but the server pauses at specific times $t'_m$ to wait for new commands before rendering the next frame, as illustrated in Fig.~\ref{fig:interactivity}. For instance, at $t'_4$ the server expects commands \{$C_1$, $C_2$\}, and at $t'_6$ it waits for $C_3$. This enforced timing ensures that every newly rendered frame accurately reflects the latest player input, maintaining the realistic cycle of action and reaction found in a real-world CG platform.

\texttt{CGReplay} interactivity faces two challenges: \textit{command loss} and \textit{frame loss}, requiring reliability mechanisms.

\begin{figure}[H]
  \centering
  %\includesvg[width=0.9\columnwidth]{4_commandloss}
  \includegraphics[width=0.9\columnwidth]{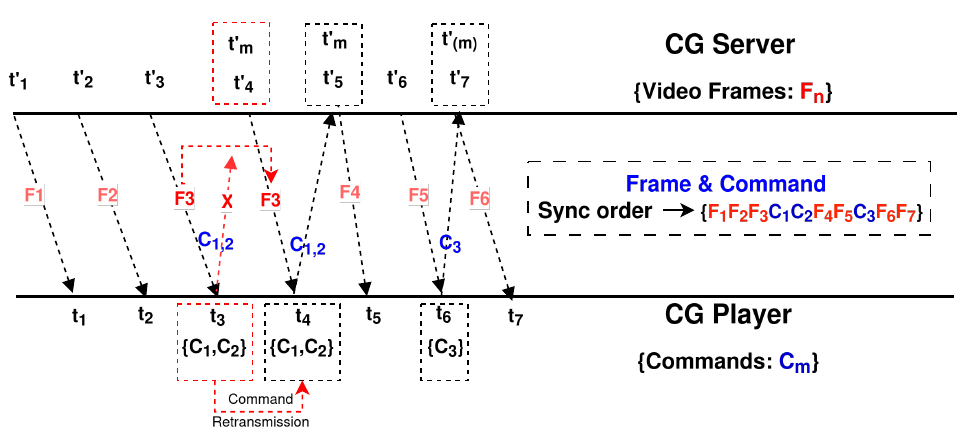} % or adjust width as needed
  \caption{Command loss scenario. \{$C_1, C_2$\} are lost through the network. Frame $f_3$ is retransmitted and the CG server waits for \{$C_1$, $C_2$\}.}
  \label{fig:cmdloss}
\end{figure}

\noindent\textbf{Command Loss Mechanism:}
%\subsubsection{Command Loss Mechanism}
command loss in CGReplay can occur due to network conditions. To address this, a recovery mechanism is implemented in both the server and player. As shown in Fig.~\ref{fig:cmdloss}, if \{$C_1$ ,$C_2$\} is lost, the CG server, while waiting for them, retransmits the previous frame to signal the player of the missing command -- or due to excessive delay. Upon detecting two consecutive frames with the same $F_{ID}$, the player re-sends the previous command. The CG server resumes sequential frame streaming once it receives the expected commands, ensuring continuity in interactivity. 

%In the \textit{command loss} scenario, the CG server expects to receive commands C2 and C3 at time $t'_m$ (corresponding to $t'_3$) to stream the next frame. However, when these commands are lost, the server retransmits the previous frame until the correct command sequence is eventually received. Once the missing commands (\textit{e.g.}, commands $C_1$ and $C_2$) arrive, the server streams the next frame (frame $f_4$) and resumes regular operation. This retransmission mechanism ensures synchronization between the server and client and introduces a lag in the game video, as the delay in command reception postpones the display of subsequent frames as shown in Fig.~\ref{fig:cmdloss}. 

%This command loss and the resulting video lag are logged on the player side, illustrating the frame retry process (see Fig.~\ref{fig:cmdloss}).
%As shown in Fig.~\ref{fig:cmdloss}, when the command is lost during the network the frame is sent again to receive the expected command, so as shown ing Fig.~\ref{fig:cmdloss} the commands $C_1$

\begin{figure}[!t]
  \centering
  %\includesvg[width=0.9\columnwidth]{5_CGframeloss}
  \includegraphics[width=0.95\columnwidth]{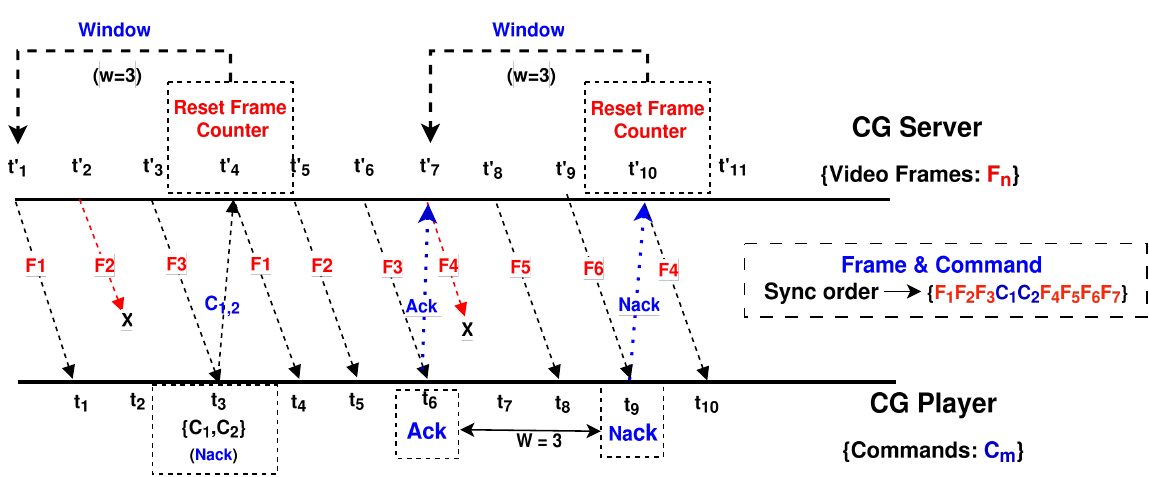} % or adjust width as needed
  \caption{Frame $f_2$ is lost while $f_3$ is received. The player notifies with NACK. ACK is sent periodically ($w=3$). NACK is sent at $t_{10}$ for frame $f_4$ loss.}
  \label{fig:frameloss}
    \vspace{-10pt}
\end{figure}

\noindent\textbf{Frame Loss Mechanism:}
to handle frame loss, we implement an ACK/NACK  control mechanism to keep the CG player synchronized with the server. ACKs and NACKs are sent in two ways: (1) as metadata with commands and (2) periodically after receiving a defined window $w$ of frames. When the player receives $w$ frames, it checks their $F_{ID}$s against the expected sync order -- sending an ACK if correct or a NACK if frames are missing or out of order. Upon receiving an ACK, the CG server continues streaming sequentially. However, if a NACK is received, the server rolls back $w$ frames and retransmits from that point. As shown in Fig.~\ref{fig:frameloss}, a NACK sent with {$C_1$, $C_2$} prompts the server to backtrack from the next expected Frame ID ($F_{ID} = 7$) to $F_{ID} = 4$, calculated as (7 - window size of 3). The server then resumes streaming from $F_{ID} = 4$, ensuring reliable frame delivery.
%As shown in Fig.~\ref{fig:frameloss}, a NACK was sent with \{$C_1$, $C_2$\}, prompting the server to back off from next expected Frame ID $F_{ID} = 7$ to the $F_{ID} = 4$, if $window = 3$, (7 - 3), and resume streaming from $F_{ID} = 4$, ensuring reliable frame delivery.
To mitigate the impact of high latency, we introduce window sliding in the CG server, allowing it to continue streaming up to $w$ frames instead of pausing at the expected frame. This increases tolerance to delays and prevents excessive interruptions, making the \texttt{CGReplay} platform smoother and more responsive.

\section{Conclusion, Demo \& Future Work}
In this work, we present \texttt{CGReplay}, an open-source platform that captures and replays user gaming sessions for CG. \texttt{CGReplay} is fully configurable through YAML files for both server and player agents and provides detailed outputs, including received video frames, frame rate, command rate, command/frame loss reports, and action/reaction response time. These features enable CG QoE evaluation under different network conditions. Additionally, \texttt{CGReplay} lays the foundation for future advancements in intelligent CG platforms and optimized encoding for CG.
%\noindent\textbf{During the demo:} participants will be able to choose between different games to be replayed. The game session is run on a single computer with a client-server topology on Mininet. While the game is being replayed, we can observe video frames and commands in addition to different QoS/QoE metrics -- \textit{e.g.}, frames per second (FPS). Furthermore, interactivity and encoding are configurable. For instance, the user may enable/disable SCReAM or even increase the target FPS. Although Mininet is used for simplicity, it can be replaced by more realistic alternatives -- \textit{e.g.}, leveraging Tofino switches.

\noindent\textbf{During the demo:} Participants can replay different games on a single computer using a client-server setup in Mininet. They can view video frames, commands, and QoS/QoE metrics like FPS. Settings such as SCReAM and target FPS are adjustable. Mininet is used for simplicity but can be replaced by realistic platforms such as Tofino switches or a more complex topology.

\section*{Acknowledgment}
This work was supported by Ericsson Telecomunicações Ltda., and by the Sao Paulo Research Foundation~(FAPESP), \fapesp~grant \texttt{2021/00199-8}, CPE SMARTNESS~\smartness.

\bibliographystyle{ieeetr}
\bibliography{Reference}
%\printbibliography

\end{document}